\documentclass [12pt] {article}

\usepackage{graphicx} 

\usepackage[a4paper, margin=2.5cm]{geometry}

\usepackage{setspace}
\usepackage{bm}
\usepackage{threeparttable}
\usepackage{xcolor}
\usepackage[T1]{fontenc}
\usepackage{titling}
\usepackage{amsmath,amssymb}
\usepackage[colorlinks=true, linkcolor=blue, citecolor=blue, urlcolor=blue]{hyperref}
\usepackage{booktabs}
\usepackage{multirow}
\usepackage{float}
\usepackage{subcaption}
\usepackage{cite}
\usepackage{comment}
\author{
\small
Felix Boakye Oppong$^{1,2}$, Dimitris Rizopoulos$^{1}$, Thierry Gorlia$^{2}$, Nicole Erler$^{3}$\\[0.5em]
\small
$^{1}$ Erasmus University Medical Center, Rotterdam, The Netherlands.\\
\small
$^{2}$ European Organisation for Research and Treatment of Cancer, Brussels, Belgium.\\
\small
$^{3}$ Julius Center for Health Sciences and Primary Care, University Medical Center Utrecht,\\ 
\small
Utrecht University, The Netherlands.\\\\ \vspace{-2em}
\small
Correspondence: \texttt{felix.oppong@eortc.org} \\ \vspace{-2em}
}
\date{}
\title{Joint modelling of time-dependent biomarker variability and time-to-event outcomes, a two-step approach. \vspace{-1em}}
\begin{document}
\maketitle
\begin{center}
\end{center}
\section*{Abstract}
Increasing evidence suggests that variability in longitudinal biomarkers, in addition to their mean trajectory, carries prognostic information for time-to-event outcomes. However, standard joint models typically capture only the expected value of the biomarker process, assuming constant residual variability across individuals and time. Fully joint extensions that model within-subject variability exist but are computationally demanding and require dedicated software packages.\\ \\
\noindent
We propose a flexible two-step approach for incorporating biomarker variability into joint models. First, residuals (or their transformations) from a mixed-effects model are used to derive subject- and time-specific measures of variability. Second, these variability measures are included in a standard joint model, allowing their association with survival to be estimated alongside the mean biomarker trajectory. Our approach can also accommodate multiple biomarkers simultaneously and is readily implemented using existing joint modeling software without custom extensions.\\ \\
\noindent
Through simulations, we show that our method provides reasonable performance for variability effects across a range of scenarios. We further illustrate our approach using longitudinal data of white blood cell counts from a large phase III glioblastoma trial, demonstrating that both mean levels and variability of hematological markers carry prognostic information for overall survival.

\newpage
\section{Introduction}
In oncology, numerous studies have investigated whether hematological parameters such as absolute neutrophil count, platelet count, hemoglobin concentration, lymphocyte count, and other biomarkers are associated with oncological outcomes such as progression-free survival (PFS) or overall survival (OS) \cite{Tanizaki2018PeripheralNivolumab, Saito2018PrognosticPatients, Schernberg2018NeutrophiliaChemoradiation, Rhun2022PrognosticGlioblastoma}. These biomarkers are typically measured repeatedly over time, and joint models for longitudinal and time-to-event outcomes provide a valuable framework to study their relationship with clinical outcomes. Joint models typically combine a mixed-effects model for the longitudinal biomarker(s) with a time-to-event model, analyzing both data types simultaneously \cite{Rizopoulos2012JointR}. In so doing, they provide more accurate estimates of how temporal changes in biomarkers affect the risk of clinical events, thereby offering a more comprehensive understanding of the often complex temporal relationship between biomarker trajectories and clinical outcomes \cite{Rizopoulos2012JointR}.\\\\
\noindent
In most applications, it is assumed that the risk of an event at any time is associated either with the estimated value of the longitudinal measurement at that time point or, more generally, with the subject-specific random effects from the longitudinal submodel \cite{Tsiatis2004JointOverview,Henderson2000JointData, Rizopoulos2012JointR, Rizopoulos2011ATime-to-event}. However, the relationship between the longitudinal measurements and the time-to-event outcome is typically more complex, as different characteristics of a longitudinal profile may be associated with the risk of an event \cite{Rizopoulos2014CombiningAveraging}. To address this, alternative functional forms (association structures) have been explored. Common choices include the time-dependent slope, which links the event risk to the rate of change of the biomarker \cite{Ye2008SemiparametricApproach, Taylor2013Real-TimeModels}, and the cumulative effect, which captures the influence of the biomarker's historical trajectory \cite{Brown2009AssessingHIV/AIDS, Rizopoulos2012JointR}, typically summarized by the area under the estimated longitudinal curve \cite{Rizopoulos2014CombiningAveraging}. \\\\
\noindent
Emerging evidence suggests that the variability in longitudinal biomarkers may hold prognostic value. In oncology and cardiology, for instance, increased instability in disease markers has been associated with adverse outcomes \cite{Piatek2020RisingCancer, Wu2023Visit-to-visitAnalysis}. This observation underpins our current investigation, in which we use data from a randomized, multicentre, phase 3 trial in patients with newly diagnosed glioblastoma to examine the hypothesis that variability in hematological parameters is likely associated with the risk of death.\\\\
\noindent
Despite the relevance in many clinical fields, there is limited research on incorporating biomarker variability as a subject-specific and time-dependent predictor in time-to-event models. Gao \textit{et al.} \cite{Gao2011AOutcome} and Barrett \textit{et al.} \cite{Barrett2019EstimatingStudy} explored this by extending the standard joint model framework to include subject-specific residual variance. However, their models adopt a restrictive association structure, where the event risk depends solely on the random effects and not on other time-varying features of the longitudinal trajectory, such as the current value or slope. Moreover, although they incorporate subject-specific residual variance through random effects, this variance is assumed to be constant over time and does not capture time-dependent changes in variability. This framework has been further extended by Martins \textit{et al.} \cite{Martins2022AHeterogeneity} to include time-varying regression coefficients, which capture the dynamic effect of a time-invariant marker variance on the event risk.
\\

\noindent
Using frequentist approaches, Li \textit{et al.} \cite{Li2023AOutcome} and Courcoul \textit{et al.} \cite{Courcoul2025AEvents} have also proposed models that allow for time-dependent subject-specific variability in the longitudinal biomarker. While the model by Li \textit{et al.}\cite{Li2023AOutcome} still restricts the association to random effects, the approach by Courcoul \textit{et al.} \cite{Courcoul2025AEvents} enables more flexible functional forms linking the biomarker and event processes. Palma \textit{et al.} \cite{Palma2025AVariability} have also proposed a model that incorporates within-subject marker variability. Their approach uses Bayesian methodology and can include up to two longitudinal biomarkers. In all of these extensions, however, dedicated software packages are required to fit these complex models, as they cannot be readily implemented using standard joint modeling software. Moreover, these specialized tools typically lack several key features of other established joint modeling packages, such as, the ability to include multiple biomarkers, handle competing risks, multi-state processes, recurrent events, dynamic prediction, etc. This limitation may hinder the widespread adoption of these advanced modeling approaches.
\\

\noindent
To address these limitations, we propose a flexible and easily implementable two-step approach for incorporating variability in longitudinal biomarkers within a standard joint modeling framework. In the first step, a mixed-effects model is fitted to the longitudinal biomarker data to obtain subject- and time-specific residuals, which reflect deviations from each subject's estimated marker trajectory. In the second step, these residuals (or their transformations) are treated as a longitudinal outcome representing biomarker variability and incorporated as time-varying covariates in a standard joint model, which can be implemented using widely available joint modeling software. Unlike previous two-stage approaches that rely on simple summary measures of variability and are not inherently time-varying (e.g. standard deviation, coefficient of variation, maximum value, etc.) \cite{Piatek2020RisingCancer, Wu2023Visit-to-visitAnalysis, Rothwell2010PrognosticHypertension}, our method makes efficient use of the longitudinal data by capturing the within-subject fluctuations over time, mitigating biases such as immortal time and regression dilution. Furthermore, because our proposed model can be fitted using standard joint modeling software, it makes full use of the features available in such packages, including handling multiple biomarkers simultaneously, competing risks, multi-state processes, recurrent events, and dynamic prediction, while remaining interpretable and computationally efficient.\\

\noindent
The structure of this paper is as follows. Section \ref{method} provides an overview of our proposed modeling approach. In Section \ref{simulation}, we present our simulation study and the simulation results. Section \ref{data} describes the motivating dataset and reports the results of the practical application of our methodology. The discussion of the paper is provided in Section \ref{discussion}.

\section{Methods}\label{method}

A standard joint model for longitudinal and time-to-event data typically combines a (multivariate) linear mixed model with a proportional hazards model to characterize the relationship between one or more repeatedly measured outcomes and a time-to-event outcome \cite{Rizopoulos2012JointR}. In this setting, the observed biomarker value for subject \(i\) at time \(t\), denoted by \(y_i(t)\), is modeled as the sum of the true subject-specific mean trajectory \(m_i(t)\) and an error term \(\varepsilon_i(t)\). The error term is commonly assumed homoscedastic (\(\varepsilon_i(t) \sim \mathcal{N}(0, \sigma^2)\)), with constant variance across all subjects and time points. However, this assumption may be overly restrictive, as individuals can differ not only in their mean trajectories but also in the variability of their biomarker measurements over time \cite{Rizopoulos2012JointR, Tsiatis2004JointOverview}. To address this, extensions to the standard model allow the residual variance to be subject-specific and time-varying \cite{Li2023AOutcome, Courcoul2025AEvents}. The longitudinal model incorporating this extension can be specified as in equation~\eqref{eq:model1}:
\begin{equation}
\begin{aligned}
y_i(t) &= m_i(t) + \varepsilon_i(t) \\
       &= \bm{x_i^\top}(t)\bm{\beta} + \bm{z_i^\top}(t)\bm{b_i} + \varepsilon_i(t), \\
       \quad \bm{b_i} &\sim \mathcal{N}( \bm{0}, \bm{\Sigma}),\quad \varepsilon_i(t) \sim \mathcal{N}(0, \sigma_i(t)^2)
\end{aligned}
\label{eq:model1}
\end{equation}

\noindent
where \bm{$\beta$} is a vector of the fixed effects parameters and \bm{$b_i$} denotes a vector of random effects, with a covariance matrix \bm{$\Sigma$}. \bm{$x_i}(t)$ and \bm{$z_i}(t)$ are the design vectors for the fixed effects \bm{$\beta$} and random effects \bm{$b_i$} respectively. $\varepsilon_i(t)$ is the measurement error, which is assumed independent of \bm{$b_i$} and has variance $\sigma_i(t)^2 $. To ensure $\sigma_i(t) $ is positive, a model may be specified for log$\{\sigma_i(t)^2\}$ or for log$\{\sigma_i(t)\}$. For instance as
\begin{equation}
\begin{aligned}
\log \{\sigma_i(t)\} &= \xi_0 + \xi_1(t) + \mu_{i0} + \mu_{i1}(t),
\end{aligned}
\end{equation}
\noindent
where $\xi_0$, $\xi_1$, $\mu_{i0}$ and $\mu_{i1}$ are the fixed and random effects  characterizing the subject-specific log-standard deviation. The corresponding time-to-event model can be defined as
\begin{equation}
h_i(t) = h_0(t) \exp\left[\bm{\gamma}^\top \bm{w_i} + \bm{\alpha_m}^\top f\{t, \bm{b}_i, \mathcal{M}_i(t)\} + \alpha_\sigma \sigma_i(t)\right],
\end{equation}
\noindent
where $\mathcal{M}_i(t) = \{m_i(u); 0 \leq u < t\}$  denotes the history of the
true unobserved longitudinal process up to time point $t$, $h_0(t)$ is the baseline hazard, $\bm{w_i}$ is a vector of baseline covariates with regression coefficients $\bm{\gamma}$, and $\bm{\alpha_m}$ is a vector of association parameters linking the longitudinal trajectory to the event risk via a function $f\{\cdot\}$ that may depend on the random effects as well as the true underlying longitudinal history  \cite{Rizopoulos2014CombiningAveraging}. The term $\alpha_\sigma$ quantifies the effect of the subject-specific time-varying standard deviation $\sigma_i(t)$, on the risk of event.

\subsection{Proposed two-step approach}

Although it is possible to explicitly model the subject-specific error variance or standard deviation within the joint modeling framework, such models are complex and often require custom programs or extensions of standard joint modeling software. Moreover, jointly estimating a full heteroscedastic variance structure alongside the longitudinal and event processes can be computationally demanding. To address these limitations, we propose a flexible and computationally efficient two-step approach that approximates the within-subject variability using residuals from a preliminary mixed-effects model.
\subsubsection*{Step 1: Estimating subject-specific residuals.} 
In the first step, we fit an initial linear mixed-effects model assuming constant residual variance. This model can be expressed as
\begin{equation*}
\begin{aligned}
y_i(t) &= m_i(t) + \varepsilon_i(t) \\
       &= \bm{x}_{1i}^\top(t)\bm{\beta} + \bm{z}_{1i}^\top(t)\bm{b}_i + \varepsilon_i(t), \\
       \quad \bm{b}_i &\sim \mathcal{N}( \bm{0}, \bm{\Sigma}), \quad \varepsilon_i(t) \sim \mathcal{N}(0, \sigma^2).
\end{aligned}
\end{equation*}
This specification represents a standard mixed-effects model for longitudinal data and can be fitted using any available software for mixed modeling, such as \texttt{lme4} or \texttt{nlme} in R \cite{Bates2015FittingLme4, Pinheiro2019Nlme:Https://cran.r-project.org/package=nlme}.

\medskip
\noindent
The fixed effects are estimated by (restricted) maximum likelihood, yielding estimates \(\hat{\bm{\beta}}\). The subject-specific random effects \(\hat{\bm{b}}_i\) are obtained using empirical Bayes estimates \cite{Pinheiro2000Mixed-EffectsS-PLUS, Verbeke2000LinearData}. Based on these estimates, the individual residuals are calculated as
\[
\hat{\varepsilon}_i(t)
=
y_i(t)
-
\bm{x}_{1i}^\top(t)\hat{\bm{\beta}}
-
\bm{z}_{1i}^\top(t)\hat{\bm{b}}_i .
\]
which represent deviations of the observed biomarker values from the fitted subject-specific trajectories. These residuals capture the within-subject fluctuations around the expected longitudinal trajectory.  

\medskip
\noindent
If the underlying biomarker trajectory is expected to be nonlinear, a nonlinear effect of time may be modeled in this step.

\subsubsection*{Step 2. Modeling biomarker variability} 
The second step defines the longitudinal submodel for the biomarker variability in the joint model, which is then linked to the time-to-event process.

\medskip
\noindent
In this step, the squared residuals \(\hat{\varepsilon}_i(t)^2\) obtained from Step 1 are used as empirical approximations of the subject- and time-specific error variance,
\[
\hat{\sigma}_i(t)^2 \approx \hat{\varepsilon}_i(t)^2,
\]
and included in the joint model as an (additional) longitudinal outcome. Analogous to how longitudinal sub-models are specified for modeling a biomarker's patient-specific mean trajectory, the squared residuals are included in the joint model using a linear mixed-effects model that captures both population-level and subject-specific patterns in the biomarker variability,
\begin{equation*}
\begin{aligned}
\hat{\varepsilon}_i(t)^2 &= R_i(t) + \eta_i(t) \\
 &= \bm{x}_{2i}^\top(t)\bm{\psi} + \bm{z}_{2i}^\top(t)\bm{u}_i + \eta_i(t),
\end{aligned}
\end{equation*}
where \(\bm{\psi}\) are the fixed effects describing the average evolution of variability, \(\bm{u}_i\) are subject-specific random effects, and \(\eta_i(t)\) is the residual error term. The random effects are assumed to follow a multivariate normal distribution, \(\bm{u}_i \sim \mathcal{N}(\bm{0}, \bm{\Omega})\), and the residual errors are assumed to be independent and normally distributed, \(\eta_i(t) \sim \mathcal{N}(0, \tau^2)\). 

\medskip
\noindent

\medskip
\noindent
An alternative is to model the absolute residuals, treating \(\hat{\sigma}_i(t) \approx |\hat{\varepsilon}_i(t)|\). In this case, the longitudinal submodel can be specified as follows:
\[
|\hat{\varepsilon}_i(t)|
= R_i(t) + \eta_i(t),
\]
with $R_i(t)$ as already defined. This formulation has practical advantages: the residuals remain on the same scale as the original biomarker, making interpretation more intuitive, and the values tend to be less extreme than the variance. However, modeling with absolute residuals or directly using the sample standard deviation introduces bias, as the sample standard deviation is a known biased estimator of the population standard deviation, particularly in small samples \cite{Gurland1971ADeviation,Brereton2015TheDistribution}. Nevertheless, this trade-off may be acceptable in exchange for computational feasibility and robustness to extreme residuals in modeling the variance.

\medskip
\noindent
Accordingly, the time-to-event submodel is specified as
\[
h_i(t) = h_0(t) \exp\left[\bm{\gamma}^\top \bm{w}_i + \bm{\alpha}_m^\top f\{t, \bm{b}_i, M_i(t)\} + \alpha_\varepsilon R_i(t)\right],
\]
where \(|\hat{\varepsilon}_i(t)|\) approximates the subject-specific marker standard deviation.\\

\noindent
This two-step approach is practical and easy to implement using standard joint modeling software, without requiring modification of existing tools. For our implementation, we employed the \texttt{JMbayes2} package in R \cite{JMbayes2}, which fits joint models under a Bayesian framework. The time-to-event submodel is specified as a proportional hazard model, assuming a multiplicative effect of covariates on the hazard function. By default, the baseline hazard \( h_0(t) \) is modeled flexibly using penalized B-splines, such that
\[
\log \{ h_0(t) \} = \sum_{p=1}^{P} \gamma_p B_p(t, \lambda),
\]
where \( B_p(t, \lambda) \) denotes the \(p\)-th B-spline basis function with knots \(\lambda_1, \ldots, \lambda_P\), and \(\bm{\gamma} = (\gamma_1, \ldots, \gamma_P)^\top\) are the corresponding spline coefficients. Posterior distributions of all model parameters, including fixed effects, random effects, and association parameters (e.g., \(\alpha_\varepsilon\)), are obtained via Markov chain Monte Carlo (MCMC) sampling under the Bayesian framework. Posterior summaries, such as means and credible intervals are used to quantify the effect of the subject-specific biomarker variability on the risk of the event.\\

\noindent
In contrast to previous two-stage approaches which summarize within-subject biomarker variability over the observation period using simple descriptive statistics such as the standard deviations \cite{Piatek2020RisingCancer, Wu2023Visit-to-visitAnalysis, Rothwell2010PrognosticHypertension}, our approach models variability as a time-varying quantity. By capturing variability at each time point, our method makes efficient use of the longitudinal data and preserves the temporal information in the biomarker trajectory. This reduces biases such as immortal time bias and regression dilution, which occur when measurements do not fully capture the true variability and can underestimate the association with event risk, particularly with sparse observations \cite{Prentice1982CovariateModel}. Furthermore, because our approach is implemented within a standard joint modeling framework, it leverages all the features of joint models, including the ability to incorporate multiple biomarkers simultaneously, flexible association structures, handling of competing risks, multi-state processes, recurrent events, dynamic prediction, etc., while remaining interpretable and computationally efficient.

\section{Simulation study}\label{simulation}
The aim of the simulation study was to evaluate the performance and robustness of the proposed two-step approach under a variety of realistic scenarios. The main simulation framework was based on Courcoul \textit{et al. }\cite{Courcoul2025AEvents}, whose design was derived from the PROGRESS clinical trial evaluating blood-pressure lowering treatment in secondary prevention \cite{MacMahon2001RandomisedAttack}. In our simulation, we considered varying levels of association between the marker variability and event risk, different sample sizes, and non-linear marker trajectories, to reflect a range of practical settings. For each simulation setup, 300 datasets were generated, and the performance of the method was assessed by comparing the estimated quantities to the true underlying values used in the data generation. A detailed description of the simulation setup is provided in the following subsections. 
\subsection{Simulating longitudinal biomarker}

Repeated biomarker measurements were simulated at 13 predefined time points specifically at $t = 0$, $0.25$, $0.5$, $0.75$, $1$, $1.5$, $2$, $2.5$, $3$, $3.5$, $4$, $4.5$, and $5$. The biomarker measurements $(y_i(t))$ were generated using a linear mixed-effects model with random intercept, random slope, and subject-specific heterogeneous error variance:
\begin{align*}
y_i(t) &= \beta_0 + \beta_1 t + b_{0i} + b_{1i} t + \varepsilon_i(t), \text{ with } \\[4pt]
\beta_0 &= 142, \quad \beta_1 = 3, \\[6pt]
\begin{pmatrix}
b_{0i} \\
b_{1i}
\end{pmatrix}
&\sim \mathcal{N}\!\left(
\begin{pmatrix}
0 \\
0
\end{pmatrix},
\boldsymbol{\Sigma}_b
\right),
\quad
\boldsymbol{\Sigma}_b =
\begin{pmatrix}
207.36 & -17.28 \\
-17.28 & 9.22
\end{pmatrix}, 
\quad \varepsilon_i(t) \sim \mathcal{N}\!\left(0, \sigma_i(t)^2\right).
\end{align*}
\noindent
The within-subject standard deviation of the biomarker, $\sigma_i(t)$, was simulated using the following log-linear model with both fixed and random effects,
$$
\text{log}\{\sigma_i(t)\} = \xi_0 + \xi_1(t) + \mu_{i0} + \mu_{i1}(t).
$$
The fixed effects were set to $\xi_0 = 2.4$ and $\xi_1 = -0.05$, and the random effects were assumed to have the following bivariate normal distribution: 
\medskip
$$
\begin{pmatrix}
\mu_{0i} \\
\mu_{1i}
\end{pmatrix}
\sim \mathcal{N} \left[
\begin{pmatrix}
0 \\
0
\end{pmatrix},
\begin{pmatrix}
0.0001 & -0.0006 \\
-0.0006 & 0.0157
\end{pmatrix}
\right]
$$
\\
\noindent 
To assess the robustness of the proposed approach for non-linear biomarker trajectories, we introduced curvature using a quadratic term centered at time 2, resulting in the following model for the longitudinal outcome: 
$$ 
y_i(t) = \beta_0 + \beta_1 t + \beta_2 (t - 2)^2 + b_{0i} + b_{1i} t + b_{2i} (t - 2)^2 + \varepsilon_i(t),
$$
\noindent where $\beta_0 = 142$, $\beta_1 = 2$, and $\beta_2 = 8$. The random effects $b_{0i}$, $b_{1i}$, and $b_{2i}$ were assumed to have the following multivariate normal distribution:
\\
$$
\medskip
\begin{pmatrix}
b_{0i} \\
b_{1i} \\
b_{2i}
\end{pmatrix}
\sim \mathcal{N} \left[
\begin{pmatrix}
0 \\
0 \\
0
\end{pmatrix},
\begin{pmatrix}
100.00 & 0 & 0 \\
0 & 9.22 & 0 \\
0 & 0 & 10.00 
\end{pmatrix}
\right] 
$$

\noindent
For the corresponding log-linear model for the within-subject standard deviation, $\sigma_i(t)$,  $\xi_0$ and $\xi_1$ were set to 2.0 and 0.08 respectively. 

\subsection{Simulation of Event Time}

Event times were simulated using the following proportional hazards model:
$$
h_i(t) = h_0(t) \exp\left(\alpha_{m} m_i(t) + \alpha_{\sigma} \sigma_i(t) \right),
$$
where \(h_0(t)\) is the baseline hazard function assumed to follow a Weibull distribution defined as \(h_0(t) = \kappa t^{\kappa - 1} e^{\zeta}\), with shape parameter \(\kappa\) and scale parameter \(\zeta\). The true underlying marker value, \(m_i(t)\), was defined according to the underlying longitudinal model 
$$
m_i(t) = \beta_0 + \beta_1 t + b_{0i} + b_{1i} t,
$$
in the linear scenario. For the model with non-linear trajectories, this was specified as
$$
m_i(t) = \beta_0 + \beta_1 t + \beta_2 (t - 2)^2 + b_{0i} + b_{1i} t + b_{2i} (t - 2)^2.
$$
\noindent
 Across all scenarios, the association between the true marker level and event risk was fixed at \(\alpha_{m} = 0.02\). To evaluate the performance of the proposed approach under varying strengths of association, the association between the marker variability and event risk, \(\alpha_{\sigma}\), was varied across 0.02, 0.07 and 0.10. To ensure comparable event rates across the different scenarios, the shape and scale parameters of the Weibull baseline hazard was varied based on the value of the association parameter \(\alpha_{\sigma}\) and whether the longitudinal model was linear or non-linear (see Table \ref{tab:weibull} in the supplementary document). Event times were generated by numerically inverting the cumulative hazard function using Brent’s root-finding method \cite{Brent1974AlgorithmsDerivatives}. Uninformative right-censoring was applied by drawing censoring times from a uniform distribution.


\subsection{Simulation results}
\begin{figure}[H]
    \centering
    \includegraphics[width=1\linewidth]{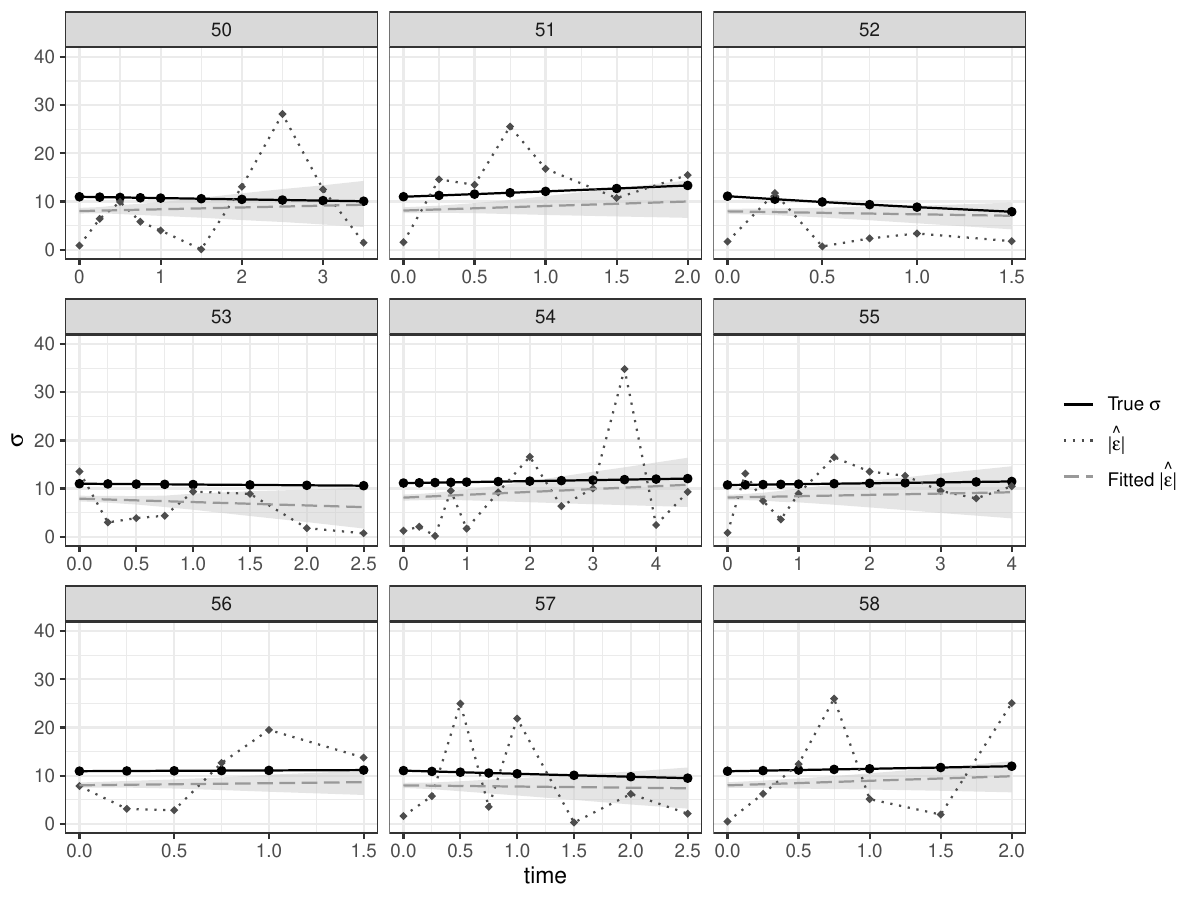}
    \caption{Simulation results with one dataset for the model with \textbf{linear marker trajectory} with $\alpha_\sigma=0.02$ and sample size of \textbf{500 subjects}}
    \label{fig:boxplot_combined}
\end{figure}

\begin{figure}[H]
    \centering
    \includegraphics[width=1\linewidth]{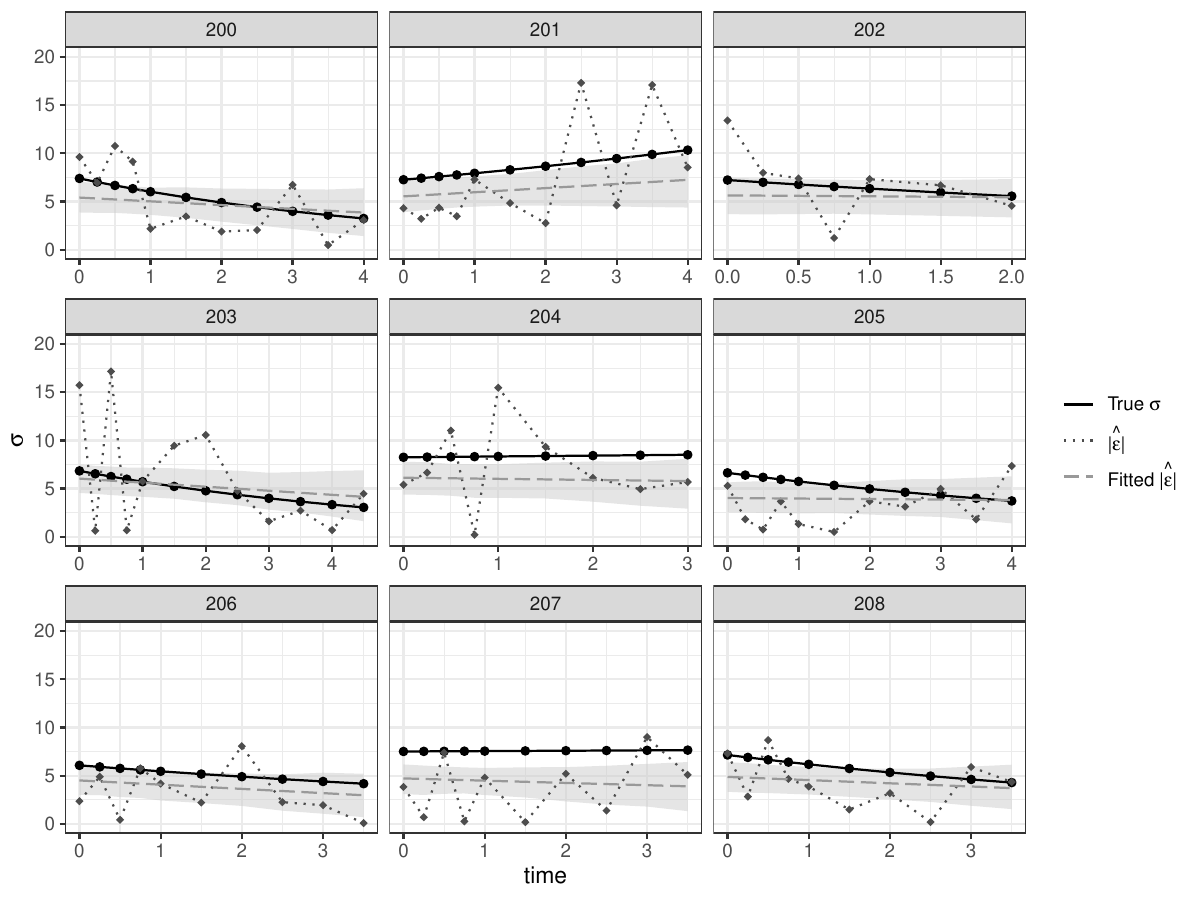}
    \caption{Simulation results with one dataset for the model with \textbf{non-linear marker trajectory} with $\alpha_\sigma=0.02$ and sample size of \textbf{500 subjects} }
    \label{fig:boxplot_combined2}
\end{figure}

The joint models were fitted using the \texttt{JMbayes2} package in \texttt{R} \cite{JMbayes2}, which employs a full Bayesian framework to estimate the longitudinal and survival submodels simultaneously via Markov Chain Monte Carlo (MCMC) sampling. Convergence of the MCMC chains was assessed using the potential scale reduction factor ($\hat{R}$), a measure that compares the within- and between-chain variances, where values close to 1 indicate good convergence. In our simulation study, parameter estimates with $\hat{R} > 1.1$ were considered not to have converged. To ensure the robustness of the reported summary statistics for $\alpha_m$ and $\alpha_\sigma$, which were more sensitive to convergence issues, estimates were excluded from the final analysis if their corresponding $\hat{R}$ exceeded this threshold. \\\\
\noindent
Table  \ref{tabsim1} reports the averages of the posterior means, the absolute bias, empirical standard errors (ESE; the standard deviation of the estimates across the 300 simulated datasets), the mean posterior standard deviations (MPSD; the average of the posterior standard deviations from each dataset), and the coverage rates of the 95\% credible intervals for the estimated parameters. These results are based on the simulation scenario with a linear marker trajectory and a sample size of 500 subjects, considering varying strengths of association between the marker variability and the event risk (\(\alpha_{\sigma} = 0.02, 0.07, 0.10\)). Additional results for the  simulations with a non-linear marker trajectory are reported in table \ref{tab:non-linear} in the Supplementary material. Simulations with a smaller sample size of 100 subjects (linear marker trajectory) encountered convergence issues within a reasonable computational time (i.e., less than 30 minutes per dataset). Consequently, results for these scenarios are not reported, as they may not reliably reflect the performance of the proposed approach. Compared to the true underlying standard deviation ($\sigma$), the fitted trajectories closely follow the true trajectory across subjects and time points, providing a good fit to the observed marker dynamics (figure \ref{fig:boxplot_combined}). Similar results are shown for the model based on non-linear marker trajectory in figure \ref{fig:boxplot_combined2}.\\\\
\noindent
Overall, our proposed approach provides satisfactory results across all simulation scenarios, with minimal bias in the parameter estimates. The mean posterior standard deviations were close to the empirical standard errors, indicating that the posterior uncertainty estimated by the model accurately reflects the variability observed across simulations. The coverage rates of the 95\% credible intervals were generally close to the nominal value. Larger differences between MPSD and ESE, and slightly reduced coverage, were observed for stronger associations (e.g., for $\alpha_{\sigma}$=0.10). 
\begin{table}[H]
\centering
\caption{Simulation results for the model with \textbf{linear marker trajectory} model across different values of $\alpha_\sigma$ with \textbf{500 subjects}}
\resizebox{\textwidth}{!}{%
\begin{tabular}{llcccccc}
\toprule
\textbf{Submodel} & \textbf{Parameter} & \textbf{True value} & \textbf{Mean} & \textbf{Bias} & \textbf{ESE} & \textbf{MPSD} &  \textbf{CR (\%)} \\
\midrule[\heavyrulewidth]
\multicolumn{8}{l}{\textbf{Scenario: $\alpha_\sigma = 0.02$}} \\
\midrule
\multirow{2}{*}{Longitudinal biomarker} 
  & $\beta_0$ (Intercept)         & 142.00   & 141.95  & -0.05 & 0.68 & 0.70 & 95.00 \\
  & $\beta_1$ (Slope)             & 3.00     & 3.02    & 0.02  & 0.23 & 0.24 & 95.33 \\
  &  &  &   &  &  &  \\
\multirow{2}{*}{Survival} 
  & $\alpha_m$ (Current value)    & 0.02  & 0.02    & 0.00  & 0.00 & 0.00 & 98.67 \\
  & $\alpha_{\sigma}$ (Current SD)& 0.02  & 0.01    & -0.01 & 0.04 & 0.03 & 93.22 \\
\midrule[\heavyrulewidth]
\multicolumn{7}{l}{\textbf{Scenario: $\alpha_\sigma = 0.07$}} \\
\midrule
\multirow{2}{*}{Longitudinal biomarker} 
  & $\beta_0$ (Intercept)         & 142.00   &     141.95  &   -0.05  &  0.67  &  0.70 &   95.00          \\
  & $\beta_1$ (Slope)             & 3.00     &   3.02     &   0.02  &   0.23 & 0.24  &     94.33   \\
&  &  &   &  &  &  \\
\multirow{2}{*}{Survival} 
 & $\alpha_m$ (Current value)    & 0.02  &   0.02  &  0.00   &  0.00 & 0.01  &   98.99       \\
  & $\alpha_{\sigma}$ (Current SD)& 0.07 &  0.10 & 0.03 &  0.06  &  0.06 & 90.74     \\
\midrule[\heavyrulewidth]
\multicolumn{7}{l}{\textbf{Scenario: $\alpha_\sigma = 0.1$}} \\
\midrule
\multirow{2}{*}{Longitudinal biomarker} 
  & $\beta_0$ (Intercept)         & 142.00   &    141.95   &    -0.05  &   0.67 & 0.70  &  94.67       \\
  & $\beta_1$ (Slope)             & 3.00     &   3.03    &   0.03    &    0.23 & 0.24 &  95.33      \\
  &  &  &   &  &  &  \\
\multirow{2}{*}{Survival} 
  & $\alpha_m$ (Current value)    & 0.02  &   0.02   &  0.00  &   0.00 & 0.01  &  97.61    \\
  & $\alpha_{\sigma}$ (Current SD)& 0.10  &   0.16  &    0.06  &   0.07 & 0.07 &   85.77   \\
\bottomrule
\end{tabular}
} 
\par\vspace{0.5em}
\smallskip
\begin{minipage}{\linewidth}
\scriptsize
The mean is the average of the parameter estimates across the 300 datasets. Bias is the difference between the mean and the true value. ESE is the empirical standard error.  MPSD is the mean of the posterior standard deviation. CR is the empirical 95\% confidence interval coverage rate. Estimates for $\alpha_m$ and $\alpha_\sigma$ were excluded from the summary if their corresponding $\hat{R}$ values exceeded 1.1, indicating non-convergence. For the scenario with $\alpha_\sigma = 0.02$, no dataset was excluded for $\alpha_m$, but 5 datasets were excluded for $\alpha_\sigma$. With  $\alpha_\sigma = 0.07$, 4 datasets were excluded for $\alpha_m$, and 30 datasets were excluded for $\alpha_\sigma$.  With  $\alpha_\sigma = 0.1$, 7 datasets were excluded for $\alpha_m$, and 47 datasets were excluded for $\alpha_\sigma$. 
\end{minipage}
\label{tabsim1}
\end{table}

\section{Motivating example}\label{data}
\begin{figure}[H]
    \centering
    \includegraphics[width=1\linewidth]{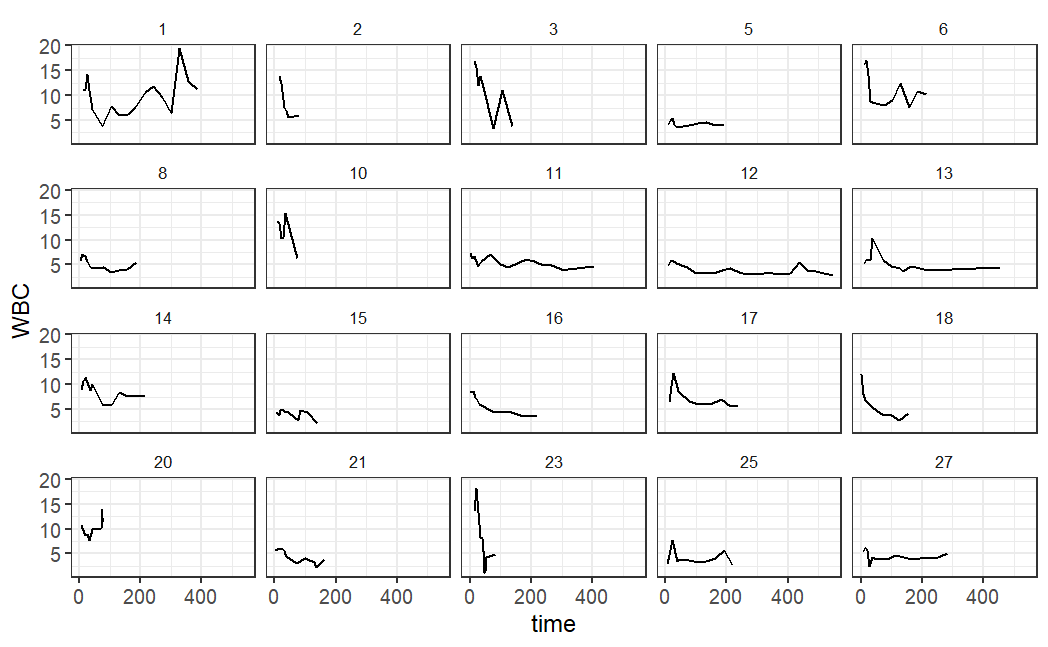}
    \caption{WBC count over time for the first 20 subjects}
    \label{fig:wbc}
\end{figure}

We illustrate our approach using data from the EORTC 1709/Canadian Cancer Trials Group CE.8 (MIRAGE) trial, a large multicentre, randomized phase 3 study \cite{Roth2024MarizomibTrial}. This trial enrolled 749 patients with newly diagnosed, histologically confirmed glioblastoma. Eligible participants were randomized 1:1 to receive standard temozolomide-based radiochemotherapy alone or in combination with marizomib, a novel pan‑proteasome inhibitor that crosses the blood–brain barrier. Treatment consisted of radiotherapy with concomitant temozolomide, followed by up to six maintenance cycles, with or without marizomib. The primary endpoint, overall survival (OS), was analyzed using Cox proportional hazards model, although the addition of marizomib to standard therapy did not improve OS (median 17.0 months vs. 16.5 months, HR = 1.04; \textit{P} = 0.64).  
\\\\
\noindent
 Hematology assessments (including hemoglobin (Hgb), hematocrit (Hct), red blood cell (RBC) count, white blood cell (WBC) count, absolute neutrophil count (ANC), platelet count) were performed weekly during the 6 weeks of radiotherapy, before the start of each maintenance cycle, and after the last temozolomide or marizomib treatment. Longitudinal variability in these hematological markers may reflect individual differences in bone marrow tolerance, systemic inflammation, or treatment response, which are not captured by baseline values alone. Examining this variability provides a clinically relevant context to investigate whether dynamic changes in hematological parameters are associated with OS, beyond established baseline factors. \\\\
\noindent
To illustrate our proposed approach, we focused on white blood cell (WBC) count, an established marker of immune function that has been linked to outcomes in patients with glioblastoma \cite{Gursoy2025CRP/albuminMultiforme, Zhang2025PeripheralTemozolomide}. The WBC measurements in this trial exhibited substantial variability both between and within patients (figure \ref{fig:wbc}), making it an appropriate marker to investigate whether dynamic changes in hematologic parameters are associated with overall survival beyond baseline values. Among the 749 randomized patients, 689 provided consent for re-use of their data for further research. After excluding 9 patients with missing WBC measurements, 680 patients were included in this illustrative analyses. The median number of repeated measurements was 10, with a range of 1-43 measurements.

\subsection{Model specification}\label{modspec}
To investigate the association between WBC variability and overall survival, we first fitted the following linear mixed model:

\begin{equation*}
\begin{aligned}
\text{WBC}_i(t) &= m_i^{\text{WBC}}(t) + \varepsilon_i(t) \\
       &=\beta_0 + \beta_1(t) + b_{0i} + b_{1i}(t) + \varepsilon_i(t), \\
       \quad \bm{b_i} &\sim \mathcal{N}( \bm{0}, \bm{\Sigma}),\quad \varepsilon_i(t) \sim \mathcal{N}(0, \sigma_i(t)^2). 
\end{aligned}
\label{eq:model33}
\end{equation*}
\noindent
The longitudinal component of the joint model then consisted of the longitudinal sub-model for WBC count, as specified above, as well as the linear mixed model for the absolute residuals (as an estimate of the  underlying subject-specific standard deviation, $\sigma_i(t)$). The model for the absolute residuals was specified as:
\begin{equation*}
\begin{aligned}
|\hat{\varepsilon}_i(t)| &= m_i^{\varepsilon}(t) +  \eta_i(t) \\
       &=\psi_0 + \psi_1 (t) + u_{0i} + u_{1i} (t) + \eta_i(t),
\end{aligned}
\end{equation*}
\noindent
where $b_{0i}, b_{1i}$ and $u_{0i}, u_{1i}$ are subject-specific random effects for the mean and absolute residuals respectively. The corresponding time-to-event submodel was specified as:
$$
\begin{aligned}
h_i(t) = h_0(t) \, \exp \big \{ 
& \gamma_{1} \, \text{Age}_i + 
  \gamma_{2} \, \text{Treatment}_i + \gamma_{3} \, \text{Surgery}_i +  \gamma_{4} \, \text{KPS}_i  + \\
& \gamma_{5} \, \text{\textit{MGMT}}_i  + 
  \alpha_m \, m_i^{\text{WBC}}(t) +  \alpha_\varepsilon \, m_i^{\varepsilon}(t) 
\big \},
\end{aligned}
$$
\noindent
where $\alpha_m$ and $\alpha_\varepsilon$ quantify the  association of the expected value and standard deviation of WBC count with the hazard of death.  The covariates included in the model are the age of the patient, the treatment assignment (TMZ/RT → TMZ (A) vs. TMZ/RT → TMZ + Marizomib (B)), type of surgery (gross total resection vs. biopsy/partial resection), KPS (Karnofsky performance status: 70/80  vs. 90/100), and \textit{MGMT} (\textit{MGMT} promoter methylation: Unmethylated vs. Methylated vs Undetermined/invalid). \\\\
\noindent
Both WBC count on its original scale and log-transformed WBC count were considered to account for skewness in the data. Moreover, given that WBC measurements may exhibit non-linear trends over time (figure \ref{fig:wbc}), joint models in which the follow-up time was modelled using natural cubic splines was fitted. The spline-based longitudinal submodel  was specified as:
$$
\text{WBC}_i(t) = \beta_0 + \sum_{k=1}^{2} \beta_k B_k(t) + b_{0i} + \sum_{k=1}^{2} b_{ki} B_k(t) + \varepsilon_i(t),
$$
\noindent
where $B_k(t)$, $k=1,2$, are the basis functions of a natural cubic spline for the time variable $t$ with 2 degrees of freedom, $\beta_k$ are the corresponding fixed-effect coefficients, and $b_{ki}$ are the subject-specific random effects associated with the spline basis functions.  This non-linear specification was used in the mixed model from which the residuals were extracted as well as in the joint model specification.

\subsection{Model Results}
The joint models described in section \ref{modspec} were fitted using the \texttt{JMBayes2} R package \cite{JMbayes2}. Model diagnostics, including inspection of residuals, were carried out for each specification. Based on these assessments, the linear model on the original WBC scale was selected to illustrate our proposed approach.

\begin{table}[ht]
\centering
\caption{Posterior means and 95\% credible intervals from the survival submodel of the joint model assessing the association of white blood cell count (WBC) with overall survival.}
\label{tab:wbc_survival}
\begin{tabular}{lccc}
\hline
Covariate & Mean & 2.5\% & 97.5\% \\
\hline
Age & 0.026 & 0.014 & 0.039 \\
Treatment (B vs. A) & 0.011 & -0.176 & 0.195 \\
Extent of Surgery (Biopsy/Partial vs. Gross total) & 0.361 & 0.183 & 0.542 \\
KPS (90/100 vs. 70/80) & -0.327 & -0.511 & -0.142 \\
\textit{MGMT} (methylated vs. Unmethylated) & -1.187 & -1.416 & -0.969 \\
\textit{MGMT} (Undetermined/invalid vs. Unmethylated) & -0.299 & -0.627 & 0.011 \\
$\alpha_m$  & -0.085 & -0.148 & -0.019 \\
$\alpha_\varepsilon$ & 0.491 & 0.068 & 0.842 \\
\hline
\end{tabular}
\begin{tablenotes}
\footnotesize
\item All parameter-specific $\hat{R}$ values were less than 1.1, indicating good convergence.
\end{tablenotes}
\end{table}
\noindent
The results as reported in table \ref{tab:wbc_survival} indicate that higher WBC count are associated with a reduced hazard of death, as reflected by a negative posterior mean for $\alpha_m$. However, greater within-subject variability in WBC count ($\alpha_\varepsilon$) shows a positive association with the risk of death, suggesting that unstable WBC profiles are detrimental for survival. These findings highlight that both the average level and the longitudinal variability in WBC count carry prognostic information beyond baseline covariates such as age, surgery, performance status, and \textit{MGMT} status.

\section{Discussion}\label{discussion}
In this work, we proposed and evaluated a two‑step approach for incorporating subject‑specific, time- dependent biomarker variability into joint models for longitudinal and time‑to‑event data. This approach was motivated by growing evidence that variability in biomarkers carries important prognostic information in many clinical settings, beyond the mean level or trajectory of the biomarker \cite{Piatek2020RisingCancer, Wu2023Visit-to-visitAnalysis}. \\\\
\noindent
Our simulation study demonstrated that our proposed method performs reasonably well across a range of scenarios, including both linear and non‑linear biomarker trajectories and for varying strengths of association between marker variability and event risk. In particular, the estimates for the association with the current biomarker value ($\alpha_m$) were unbiased and showed good coverage in all settings. Also, the estimates for the variability effect ($\alpha_\sigma$) showed good performance for weak to moderate associations but exhibited some  bias and reduced coverage as the true association became stronger. This behavior likely reflects a combination of model misspecification, the inherent challenges in estimating time‑varying variance from residuals, and the fact that the sample standard deviation is a biased estimator of the population standard deviation \cite{Brereton2015TheDistribution}. Specifically, in our simulation setup, for stronger variability effects, the use of residual-based estimates appeared to overestimate the effect.\\ 

\noindent
A key practical advantage of our two‑step approach is its ease of implementation. By deriving residual-based variability measures from a standard mixed-effects model, these measures can be incorporated into any existing joint modeling software without requiring complex extensions or custom routines. This makes the approach accessible to a wide range of users. However, this convenience comes with important limitations. Modeling absolute residuals with a linear mixed-effects model can produce negative fitted values for quantities that are inherently positive, implying that the resulting estimates may not accurately reflect the true underlying variability. Furthermore, while residuals provide a useful approximation for subject-specific variability, they are themselves estimated quantities and therefore introduce sampling variability and potential bias, particularly in small samples or when measurements are sparse  \cite{Henderson2000JointData, Guo2004SeparatePackages}. \\\\
\noindent
Care must also be taken in specifying the first-stage mixed model which is used to compute the residuals. If this model is made overly flexible, for example, by including complex spline terms in the random effects, most of the observed variation in the biomarker may be absorbed by the model, leaving little residual variation to quantify as subject-specific variability. In our motivating example, using a more flexible spline-based model to capture the trajectory of WBC count resulted in a larger estimated association with the current biomarker value but a smaller estimated effect for the variability. This illustrates that when the model explains more of the structured trajectory, less residual variation is available for estimating subject-specific variability, which can attenuate variability effects. Importantly, distinguishing between short and long-term variability typically requires a substantially larger number of repeated measurements per subject. With limited data, overly flexible models may fail to distinguish these sources of variability or fail to capture meaningful within-patient fluctuations. This underscores the importance of carefully balancing model flexibility with the available data to ensure that both structured trajectories and biologically relevant variability are adequately represented without overfitting. \\\\
\noindent
The analysis of our motivating example showed that both the mean levels and variability of hematological markers can carry prognostic information. In particular, higher subject-specific variability in WBC was associated with increased risk of death.  Using our two-step approach with \texttt{JMBayes2}, we obtained estimates that were largely consistent with those from the more complex frequentist approach of  Courcoul \textit{et al.} \cite{Courcoul2025AEvents}, who developed the \texttt{FlexVarJM} R package. This R package estimates joint models that incorporate time-dependent, subject-specific residual variance, allowing for flexible functional forms and competing events. Notably, our approach produced comparable inference for baseline covariates and variability effects, while substantially reducing computation time, (8.6 minutes with \texttt{JMBayes2} compared with over 3.6 hours using \texttt{FlexVarJM}), without compromising convergence or reliability of the estimates. This demonstrates that our two-step approach can provide efficient and interpretable inference for both mean and variability effects in longitudinal biomarker data, offering a practical alternative to fully joint, high-dimensional modeling.\\\\
\noindent
An additional practical advantage of our two-step approach is the ability to include multiple biomarkers simultaneously, incorporating both the mean and variance of each biomarker. For example, we implemented our method using the \texttt{JMBayes2} R package by including both the mean and variance of WBC count and Platelet count. In contrast, to the best of our knowledge, the model by Courcoul \textit{et al.} \cite{Courcoul2025AEvents} does not allow multiple biomarkers. In an attempt to fit their model to Platelet count, as we had done separately for WBC count, there were issues with convergence and the model could not be fitted. In comparison, the method by Palma \textit{et al. }\cite{Palma2025AVariability} allows up to only two biomarkers. This capability of our framework enhances its applicability to complex clinical scenarios involving multiple correlated biomarkers. \\\\
\noindent
Finally, while our approach offers several practical advantages, it also shares a common limitation of all two-stage methods. It does not account for the uncertainty introduced in the first-stage estimation when fitting the second-stage survival model. This can lead to underestimated variability of the association parameters and overly precise inference compared with fully joint modeling approaches, which simultaneously estimate all model components and properly account for this uncertainty \cite{Henderson2000JointData, Guo2004SeparatePackages, Rizopoulos2012JointR}.\\\\
\noindent
In summary, our two-step approach offers a practical and computationally efficient strategy for incorporating subject-specific biomarker variability into joint models, enabling broader application in clinical research without the need for complex model extensions.
Although it cannot guarantee unbiased estimates, it provides a valuable alternative for settings where flexibility and scalability are priorities.

\subsubsection*{Acknowledgments}

This work is supported by the Belgian National Lottery and their players.

\newpage

\bibliography{JM.bib, ref.bib}

\newpage
\section{Supplementary Materials}\label{Supplementary}

\begin{table}[ht]
\centering
\caption{Shape $\kappa$ and scale $\zeta$ parameters used in the simulation under different $\alpha_\sigma$ values for the model with linear and non-linear biomarker trajectory.}
\begin{tabular}{lccc}
\toprule
& \multicolumn{3}{c}{\textbf{$\alpha_\sigma$}} \\
\cmidrule(lr){2-4}
& \textbf{0.02} & \textbf{0.07} & \textbf{0.10} \\
\midrule
\multicolumn{4}{l}{\textit{Linear model}} \\
Shape ($\kappa$) & $1.8^2$ & $1.7^2$ & $1.6^2$ \\
Scale ($\zeta$) & $-7$ & $-7$ & $-7$ \\
\addlinespace
\multicolumn{4}{l}{\textit{Non-linear model}} \\
Shape ($\kappa$) & $1.8^2$ & $1.6^2$ & $1.7^2$ \\
Scale ($\zeta$) & $-8$ & $-7.5$ & $-7.5$ \\
\bottomrule
\end{tabular}

\label{tab:weibull}
\end{table}

\begin{table}[H]
\centering
\caption{Simulation results for the model with \textbf{non-linear marker trajectory}  across different values of $\alpha_\sigma$ with \textbf{500 subjects}}
\resizebox{\textwidth}{!}{%
\begin{tabular}{llcccccc}
\toprule
\textbf{Submodel} & \textbf{Parameter} & \textbf{True value} & \textbf{Mean} & \textbf{Bias} & \textbf{ESE} & \textbf{MPSD } & \textbf{CR (\%)} \\
\midrule[\heavyrulewidth]
\multicolumn{8}{l}{\textbf{Scenario: $\alpha_\sigma = 0.02$}} \\
\midrule
\multirow{3}{*}{Longitudinal biomarker} 
  & $\beta_0$ (Intercept)         & 142.00   & 142.01  & 0.01  &  0.49 &  0.53  & 96.00 \\
  & $\beta_1$ (Slope)             & 2.00     & 2.01  & 0.01   &   0.19 & 0.18  &  93.67 \\
  & $\beta_2$ (Quadratic term)    & 8.00     &   8.00  &  0.00 & 0.16 & 0.17  & 95.33 \\ \\
\multirow{2}{*}{Survival} 
  & $\alpha_m$ (Current value)    & 0.02  &   0.02   &  0.00  &  0.00  & 0.00 & 98.33 \\
  & $\alpha_{\sigma}$ (Current SD)& 0.02  &  0.02  &  0.00  & 0.04 & 0.04 & 93.98 \\
\midrule[\heavyrulewidth]
\multicolumn{7}{l}{\textbf{Scenario: $\alpha_\sigma = 0.07$}} \\
\midrule
\multirow{3}{*}{Longitudinal biomarker} 
  & $\beta_0$ (Intercept)         & 142.00   & 142.01 & 0.01 & 0.49 & 0.52 &  96.67 \\
  & $\beta_1$ (Slope)             & 2.00 &   2.01 & 0.01 & 0.19 & 0.18 & 94.67   \\
  & $\beta_2$ (Quadratic term)    & 8.00 & 8.00 & 0.00 & 0.16 & 0.17 &  94.67    \\ \\
\multirow{2}{*}{Survival} 
  & $\alpha_m$ (Current value)    & 0.02  & 0.02 & 0.00 & 0.00 & 0.00 & 98.66 \\
  & $\alpha_{\sigma}$ (Current SD)& 0.07  & 0.10 & 0.03 & 0.05 & 0.05 & 90.27 \\
\midrule[\heavyrulewidth]
\multicolumn{7}{l}{\textbf{Scenario: $\alpha_\sigma = 0.1$}} \\
\midrule
\multirow{3}{*}{Longitudinal biomarker} 
  & $\beta_0$ (Intercept)         & 142.00 & 142.00 & 0.00 & 0.53 & 0.57 & 96.67 \\
  & $\beta_1$ (Slope) & 2.00 &  2.02 & 0.02  &  0.22 & 0.22 & 93.67  \\
  & $\beta_2$ (Quadratic term)    & 8.00  &  8.01  & 0.01 & 0.18 & 0.18 & 95.00  \\ \\
\multirow{2}{*}{Survival} 
  & $\alpha_m$ (Current value)    & 0.02  &   0.02 &  0.00  & 0.00 & 0.00 & 98.98 \\
  & $\alpha_{\sigma}$ (Current SD)& 0.10  &   0.15  & 0.05 & 0.06 & 0.07 & 89.05  \\
\bottomrule
\end{tabular}
} 
\par\vspace{0.5em}
\smallskip
\begin{minipage}{\linewidth}
\small \scriptsize
For $\alpha_\sigma = 0.02$, no dataset was excluded for $\alpha_m$, but 1 datasets was excluded for $\alpha_\sigma$. With  $\alpha_\sigma = 0.07$, 1 dataset was excluded for $\alpha_m$, and 2 datasets were excluded for $\alpha_\sigma$.  With  $\alpha_\sigma = 0.1$, 6 datasets were excluded for $\alpha_m$, and 26 datasets were excluded for $\alpha_\sigma$. 
\end{minipage}
\label{tab:non-linear}
\end{table}

\end{document}